\documentclass[amsmath,amssymb,a4paper,preprint,prl]{revtex4}
\usepackage{amssymb}
\usepackage{txfonts}
\usepackage{graphicx}
\usepackage{ulem}
\usepackage{color}

\begin{document}

\title{Intrinsic magnetic topological insulators in van der Waals layered MnBi$_2$Te$_4$-family materials}

\author{Jiaheng \surname{Li}$^{1,2}$}
\author{Yang \surname{Li}$^{1,2}$}
\author{Shiqiao \surname{Du}$^{1,2}$}
\author{Zun \surname{Wang}$^{1,2}$}
\author{Bing-Lin \surname{Gu}$^{1,2,3}$}
\author{Shou-Cheng \surname{Zhang}$^{4}$}
\author{Ke \surname{He}$^{1,2}$}
\email{kehe@tsinghua.edu.cn}
\author{Wenhui \surname{Duan}$^{1,2,3}$}
\email{dwh@phys.tsinghua.edu.cn}
\author{Yong \surname{Xu}$^{1,2,5}$}
\email{yongxu@mail.tsinghua.edu.cn}

\affiliation{
$^1$State Key Laboratory of Low Dimensional Quantum Physics, Department of Physics, Tsinghua University, Beijing 100084, People's Republic of China \\
$^2$Collaborative Innovation Center of Quantum Matter, Beijing 100084, People's Republic of China \\
$^3$Institute for Advanced Study, Tsinghua University, Beijing 100084, People's Republic of China \\
$^4$Department of Physics, McCullough Building, Stanford University, Stanford, California 94305-4045, USA \\
$^5$RIKEN Center for Emergent Matter Science (CEMS), Wako, Saitama 351-0198, Japan
}


\begin{abstract}
\textbf{The interplay of magnetism and topology is a key research subject in condensed matter physics and material science, which offers great opportunities to explore emerging new physics, like the quantum anomalous Hall (QAH) effect, axion electrodynamics and Majorana fermions. However, these exotic physical effects have rarely been realized in experiment, due to the lacking of suitable working materials. Here we predict that van der Waals layered MnBi$_2$Te$_4$-family materials show two-dimensional (2D) ferromagnetism in the single layer and three-dimensional (3D) $A$-type antiferromagnetism in the bulk, which could serve as a next-generation material platform for the state-of-art research. Remarkably, we predict extremely rich topological quantum effects with outstanding features in an experimentally available material MnBi$_2$Te$_4$, including a 3D antiferromagnetic topological insulator with the long-sought topological axion states, the type-II magnetic Weyl semimetal (WSM) with simply one pair of Weyl points, and the high-temperature intrinsic QAH effect. These striking predictions, if proved experimentally, could profoundly transform future research and technology of topological quantum physics.}
\end{abstract}

\maketitle

The correlation between topology and symmetry as a central fundamental problem of modern physics has attracted intensive research interests in condensed matter physics and material science since the discovery of topological insulators (TIs)~\cite{hasan2010,qi2011}. One essential symmetry is the time reversal symmetry (TRS), which is crucial to many kinds of topological quantum states of matter like TIs and is broken in the presence of magnetism. Intriguingly, the interplay between magnetism and topology in materials could generate a variety of exotic topological quantum states~\cite{hasan2010,qi2011,armitage2018}, including the quantum anomalous Hall (QAH) effect showing dissipationless chiral edge states~\cite{haldane1988,yu2010,chang2013}, the topological axion states displaying quantized magnetoelectric effects~\cite{qi2008,wilczek1987,essin2009}, and Majorana fermions obeying non-abelian statistics~\cite{qi2011,fu2008,he2017}. In this context, great research effort has been devoted to explore the novel topological quantum physics, which is of profound importance to fundamental science and future technologies, like dissipationless topological electronics and topological quantum computation~\cite{hasan2010,qi2011}.

One key subject that is of crucial significance to the whole research community is to develop topological quantum materials (TQMs) showing the coexistence of topology and other quantum phases (e.g. magnetism, ferroelectricity, charge density wave, and superconductivity), coined ``composite TQMs'' (CTQMs), which include magnetic TQMs (MTQMs) as an important subset. Currently, very limited number of MTQMs are experimentally available, including magnetically doped TIs and magnetic topological heterostructures~\cite{chang2013,chang2015,katmis2016,mogi2017}, whose material fabrication, experimental measurement and property optimization are quite challenging. Since the major topological properties of these existing MTQMs depend sensitively on delicate magnetic doping or proximity effects, only little preliminary experimental progress has been achieved until now, leaving many important physical effects not ready for practical use or still unproved. For instance, the QAH effect was only observed in magnetically doped (Bi$_{x}$Sb$_{1-x}$)$_2$Te$_3$ thin films at very low working temperatures by fine tuning chemical compositions~\cite{chang2013,chang2015}. While a previous theory predicted the existence of topological axion states at the surface of a three-dimensional (3D) antiferromagnetic (AFM) TI~\cite{mong2010}, no 3D AFM TI has been realized experimentally, as far as we know. For future research and applications, people should go beyond the existing strategy for building MTQMs and try to design intrinsic MTQMs in no need of introducing alloy/doping or heterostructures.

Noticeably, van der Waals (vdW) layered materials represent a large family of materials with greatly tunable properties by quantum size effects or vdW heterojunctions~\cite{geim2013}, in which a variety of quantum phases in different spatial dimensions have been discovered by the state-of-the-art research~\cite{zhang2009,chen2009,qian2014,tang2017,wu2018,soluyanov2015}. For instance, both two-dimensional (2D) and 3D TI states were previously found in the tetradymite-type Bi$_2$Te$_3$-class materials~\cite{zhang2009,chen2009}, and 2D intrinsic magnetism was recently found in ultrathin films of CrI$_3$~\cite{huang2018} and Cr$_2$Ge$_2$Te$_6$~\cite{gong2017} . However, these topological materials are not magnetic; those magnetic material are not topological. It is highly desirable to incorporate the magnetic and topological states together into the same vdW material, so as to obtain layered intrinsic MTQMs, which are able to inherit advantages of vdW-family materials.

In this work, based on first-principles (FP) calculations, we find a series of layered intrinsic MTQMs from the tetradymite-type MnBi$_2$Te$_4$-related ternary chalcogenides (MB$_2$T$_4$: M = transition-metal or rare-earth element, B = Bi or Sb, T = Te, Se or S), in which the intralayer exchange coupling is ferromagnetic (FM), giving 2D ferromagnetism in their septuple layer (SL); while the interlayer exchange coupling is antiferromagnetic, forming 3D $A$-type antiferromagnetism in their vdW layered bulk. Strikingly, we predict that plenty of novel 2D and 3D intrinsic magnetic topological states with outstanding features could be naturally realized in an experimentally available material MnBi$_2$Te$_4$, including the 3D AFM TI in the bulk together with the long-sought topological axion states on the (001) surface, large-gap intrinsic QAH insulators ($E_{\textit{g}} \sim$ 38 meV) with dissipaltionless and patternable chiral edge states in thin films, the type-II magnetic WSM with a single pair of Weyl points in the FM bulk, and possibly Majorana fermions if interacting with superconductivity.

The tetradymite-type MnBi$_2$Te$_4$ crystalizes in a rhombohedral layered structure with the space group $R\overline{3}m$, and each layer has a triangular lattice with atoms ABC-stacked along the out-of-plane direction, the same as Bi$_2$Te$_3$. Slightly differently, one layer MnBi$_2$Te$_4$ includes seven atoms in a unit cell, forming a Te-Bi-Te-Mn-Te-Bi-Te SL, which can be viewed as intercalating a Mn-Te bilayer into the center of a Bi$_2$Te$_3$ quintuple layer (Fig. 1a). Noticeably, the mixing between Mn and Bi in this compound would create unstable valence states Mn$^{+3}$ and Bi$^{+2}$, which is energetically unfavorable. Thus the formation of alloys could be avoided, leading to stoichiometric compounds. This physical mechanism also explains the stability of compounds like SnBi$_2$Te$_4$ and PbBi$_2$Te$_4$ and is applicable to many other MB$_2$T$_4$-family materials. Importantly, our FP calculations predicted that a series of  MB$_2$T$_4$ compounds are energetically and dynamically stable, including M = Ti, V, Mn, Ni, Eu, etc., as confirmed by FP total-energy and phonon calculations~\cite{SM}, which could be fabricated by experiment. These stable compounds are characterized by an insulating band gap in their single layers, offering a collection of layered intrinsic magnetic (topological) insulators.

In fact, MnBi$_2$Te$_4$ is now experimentally available~\cite{lee2013}, which is selected to introduce properties of the material family. Mn has a valence charge of $+$2 by losing its two 4$s$ electrons. The remaining five 3$d$ electrons fill up the spin-up Mn-$d$ levels according to the Hund's rule, introducing 5 $\mu_{B}$ magnetic moment (mostly from Mn) per unit cell. By comparing different magnetic structures, we found that the magnetic ground state is a 2D ferromagnetism with an out-of-plane easy axis in the single layer (depicted Fig. 1a), in agreement with the previous work~\cite{otrokov2017}. Each Mn atom is bonded with six neighboring Te atoms, which form a slightly distorted edge-sharing octahedron. According to the Goodenough-Kanamori rule, the superexchange interactions between Mn-Te-Mn with a bonding angle of $\sim94^{\circ}$ are ferromagnetic, similar as in CrI$_3$ and Cr$_2$Ge$_2$Te$_6$~\cite{huang2018,gong2017}. Intralayer ferromagnetic exchange couplings was also found in other stable MB$_2$T$_4$-family members, but the easy axis can be varied to in-plane (e.g. for M = V), displaying rich 2D magnetic features.

Intriguingly, magnetic and topological states are well incorporated together into MnBi$_2$Te$_4$, where Mn introduces magnetism and the Bi-Te layers could generate topological states similar as Bi$_2$Te$_3$~\cite{zhang2009}, as schematically depicted in Fig. 1b. The exchange splitting between spin-up and spin-down Mn $d$ bands are extremely large ($>$ 7 eV) caused by the large magnetic moment of Mn. Thus Mn $d$-bands are far away from the band gap, and only Bi/Te $p$-bands are close to the Fermi level. The single layer has a 0.73 eV direct (1.36 eV indirect) band gap when including (excluding) spin-orbit coupling (SOC) (Figs. 1c and 1d). While the single layer is topologically trivial ferromagnetic insulator, extremely interesting topological quantum physics emerges in the bulk and thin films, as we will demonstrate.

For the layered bulk, we found that the interlayer magnetic coupling is AFM, giving a $A$-type AFM ground state (depicted in Fig. 2a)~\cite{SM}, which is similar as in Cr$_2$Ge$_2$Te$_6$~\cite{gong2017} and explained by the interlayer super-superexchange coupling. The spatial inversion symmetry $P$ (i.e. $P_1$ centered at O$_1$) is preserved, but the TRS $\Theta$ gets broken. There exist two new symmetries: $P_2 \Theta$ and $S=\Theta T_{1/2}$, where $P_2$ is an inversion operation centered at O$_2$ and $T_{1/2}$ is a lattice translation (depicted in Fig. 2a). The band structures with and without SOC are presented in Figs. 2b and 2c. Here every band is (at least) doubly degenerate, which is ensured by the $P_2 \Theta$ symmetry~\cite{tang2016}. In the presence of $S$, a $\mathcal{Z}_2$ classification becomes feasible~\cite{mong2010}. However, in contrast to the time-reversal invariant case, there is a $\mathcal{Z}_2$ invariant for the $k_z = 0$ plane but not for the $k_z = \pi$ plane. $\mathcal{Z}_2 = 1$ corresponds to a novel 3D AFM TI phase~\cite{mong2010}, which has not been experimentally confirmed.

Here the parity criteria can be applied to determine $\mathcal{Z}_2$ due to the $P_1$ symmetry~\cite{fu2007}. Interestingly, the parities of the valence band maximum (VBM) and conduction band minimum (CBM) at $\Gamma$ are opposite, and both change signs by the SOC effects (Figs. 2b and 2c), implying a band inversion. By varying the SOC strength, the band gap first closes and then reopens at $\Gamma$ (Fig. 2d), showing simply one band inversion and thus implying a topological phase transition. The SOC-induced band reversal is happened between Bi $p_z^+$ and Te $p_z^-$, essentially the same as for Bi$_2$Te$_3$~\cite{zhang2009}. Our Wannier charge center (WCC) calculations for the $k_z = 0$ plane revealed $\mathcal{Z}_2 = 1$ ($\mathcal{Z}_2 = 0$) for with (without) SOC, confirming that MnBi$_2$Te$_4$ is a 3D AFM TI. The global band gap is $\sim$0.16 eV direct at $Z$, and the direct gap at $\Gamma$ is $\sim$0.18 eV. One prominent feature of the AFM TI is the existence of 2D gapless surface states protected by $S$, which is confirmed by the surface-state calculations (Figs. 2f and 2g). The surface states are indeed gapless on the (100) termination. However, they become gapped on the (001) termination due to the $S$ symmetry breaking.

The intrinsically gapped (001) surfaces are promising for probing the long-sought topological axion states, which give the topological quantized magnetoelectric effect related to an axion field with $\theta = \pi$~\cite{qi2008,wilczek1987,essin2009,mong2010}. Previous work proposed to probe the novel states by adding opposite out-of-plane ferromagnetism onto the bottom and top surfaces of 3D TIs~\cite{qi2008,mogi2017}. Such kind of TRS-broken surface states are naturally provided by even-layer MnBi$_2$Te$_4$ films (when with negligible hybridizations between top and bottom surface states), benefitting from their $A$-type AFM structure. An additional requirement is to open a band gap at side surfaces, which is realized in relatively thin films or by breaking the $S$ symmetry on the side surfaces (e.g. controlling surface morphology). The simplified proposal together with the suitable material candidate MnBi$_2$Te$_4$ could greatly facilitate the research of axion electrodynamics.

The AFM ground state of MnBi$_2$Te$_4$ could be tuned, for instance, by applying an external magnetic field, to other magnetic structures. Then the material symmetry changes, leading to distinct topological phases. This concept is demonstrated by studying the simple FM ordering along the out-of-plane direction (Fig. 3a). The FM structure has $P_1$ but neither $\Theta$ nor $P_2 \Theta$, leading to spin-split bands, nonzero Berry curvatures and possibly nonzero topological Chern numbers that correspond to novel topological phases, like 3D QAH insulators~\cite{jin2018} and WSMs~\cite{wan2011,weng2015,xu2015,soluyanov2015,armitage2018}. The band structure of FM MnBi$_2$Te$_4$ (Fig. 3b) displays a pair of band crossings at W/W$'$ along the $\overline{\textit{Z}}$-$\Gamma$-Z line. The band crossings are induced by interlayer orbital hybridizations and protected by the $C_3$ rotational symmetry. Our WCC calculations found that W is a momentum-space monopole with a topological charge of +1 (i.e. an Berry phase of $2\pi$) (Fig. 3e), and its time-reversal partner W$'$ has an opposite topological charge of -1, indicating that the system is a topological WSM. While in most of the momentum space the electron pocket is located above the hole pocket (e.g. along the F-W-L line), the Weyl cones get tilted along the $\overline{\textit{Z}}$-$\Gamma$-Z line (Fig. 3d), leaving some part of the electron pocket below the hole pocket, which is the characteristic feature of the type-II WSM~\cite{soluyanov2015}. In contrast to the time-reversal-invariant WSMs that must have even pairs of Weyl points, this ferromagnetic WSM represents the simplest one, hosting only a pair of Weyl points. Moreover, our surface-state calculations clearly demonstrated the existence of Fermi arcs on the (100) and ($\bar{1}$00) terminations (Figs. 3f and 3g), which is the fingerprint of the WSM. Importantly, the Weyl points are well separated in the momentum space and very close to the Fermi level, advantageous for experimental observations.

The vdW layered materials are featured by tunable quantum size effects. For AFM MnBi$_2$Te$_4$ films, even layers do not possess $P$ and $\Theta$ symmetries, but have $P_2 \Theta$, ensuring double degeneracy in every band. Differently, odd layers have $P$ but neither $\Theta$ nor $P_2 \Theta$, leading to spin-split bands. The different symmetries lead to distinct topological properties in even and odd layers. Specifically, the topological Chern number $\mathcal{C} = 0$ is required by $P_2 \Theta$ in even layers, $\mathcal{C} \ne 0$ is allowed in odd layers. In contrast to the single and 3-layer films, where a trivial insulating gap is opened by quantum confinement effects, we found that the 5-layer film is an intrinsic QAH insulator with $\mathcal{C} = 1 $, as confirmed by the appearance of an quantized Hall conductance and chiral edge states within the bulk gap (Figs. 4a-c). We also calculated a 7-layer film, which is also a QAH insulator with $\mathcal{C} = 1 $.

The exchange splitting introduced by the magnetic Mn-layers together with the strong SOC effects in the Bi-Te bands cooperatively induce the QAH effect, similar as in the magnetically doped Bi$_2$Te$_3$-class TIs~\cite{yu2010}. However, the present material has an intrinsic magnetism and does not need uncommon mechanisms (e.g. the Van-Vleck mechanism~\cite{yu2010}) to form ferromagnetism in insulating states. Moreover, since the magnetic doping is not needed, disorder-induced magnetic domains and potential fluctuations, that deteriorate the QAH effect, are avoided. Furthermore,  the QAH gap of the 5-layer film is 38 meV, greater than the room-temperature thermal energy of 26 meV, enabling a high working temperature.

The thickness dependence behaviors can be understood as follows. There are intrinsically gapped surface states on both sides of thick films, as obtained for the (001) semi-infinite surface. These surface gaps are opened by the TRS-breaking field near the surface, which have half quantized Hall conductances $\sigma_{xy} = e^2/2h$ or $-e^2/2h$ when the magnetism in the surface layer is up- or down-oriented, respectively. Thus, the Hall conductances of bottom and top surfaces is cancelled in even layers, giving axion insulators with $\mathcal{C} = 0$ or topological axion states mentioned above; while they get added in odd layers, giving QAH insulators with $\mathcal{C} = 1$. This physical picture, consistent with our calculation results, suggests an oscillation of $\mathcal{C}$ in even and odd layers. Thus chiral edge states always appear near step edges (Fig. 4d), which can be used for dissipationless conduction. Based on this unique feature, the chiral edge states could be selectively patterned by controlling the film thickness or step edges, advantageous for building dissipationless circuits.

Looking back to the history of the TI research, the first- and second-generation TIs are the HgTe/CdTe quantum wells~\cite{bernevig2006} and Bi-Sb alloys, respectively, which are very complex and difficult to study theoretically and experimentally. Research interests have been increased exponentially since the discovery of the third-generation TIs in the intrinsic Bi$_2$Te$_3$-class materials~\cite{zhang2009,chen2009}. A very similar situation is faced by the research of magnetic topological physics. Currently, experimental works are majorally based on magnetically doped TIs and magnetic topological heterostructures, which are quite challenging and have led to little preliminary progress. Looking forwards, the research progress is expected to be greatly prompted by discovering intrinsic MTQMs that are simple and easy to control. The vdW layered MnBi$_2$Te$_4$-family materials satisfy all these material traits. More importantly, this material family could host extremely rich topological quantum states in different spatial dimensions (like 3D AFM TIs, time-reversal invariant TIs and topological semimetals, 2D QAH and QSH insulators, etc.) and are promising for investigating other exotic emerging physics (like Majorana fermions), which are thus perfect next-generation MTQMs for future research.


\begin{figure}
\includegraphics[width=\linewidth]{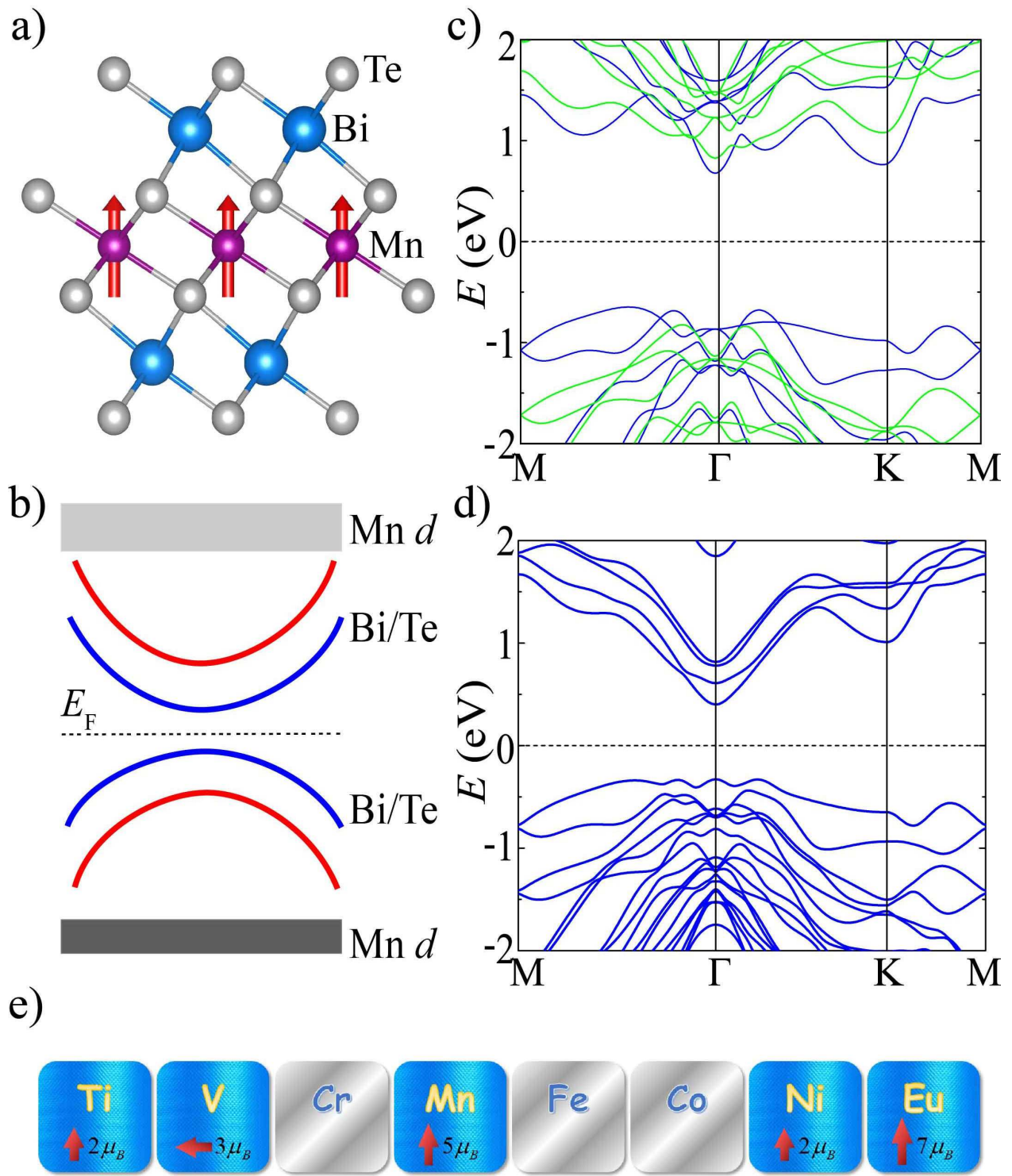}
\caption{Monolayer MnBi$_2$Te$_4$ (MBT). (a) The side view of monolayer MBT. (b) A schematic diagram of band structure in MBT including thin films and 3D bulk. (c,d) Band structures of monolayer MBT calculated by HSE06 without/with spin-orbital coupling (SOC). The blue (green) curve in the (1c) represents the spin-up (spin-down) states. (e) A schematic diagram of XBT materials family. The red arrows represent the magnetic moment, whose length and direction represent the magnitude and easy magnetic axis of magnetic atoms.}
\end{figure}

\begin{figure}
\includegraphics[width=\linewidth]{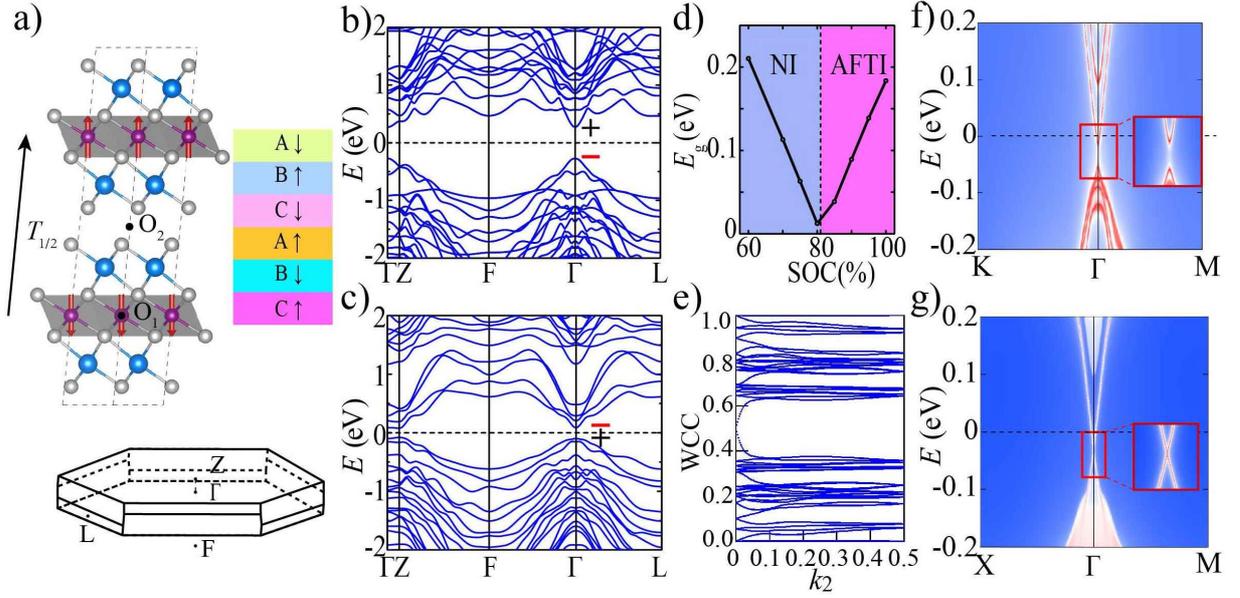}
\caption{AFM MBT. (a) The side view of bulk AFM MBT crystal structure.The black dot O$_1$ represents inversion center located at the magnetic Mn atoms. The black dot O$_2$ represents symmetry operation which combines inversion operator and the lattice translation $T_{1/2}$  along c axis. The schematic diagram shows MBT is stacked in the way combining the AFM order and ABC-stacking sequence. Brillouin zone used in the calculation are showed below. (b), (c) Band structures of 3D bulk AFM MBT without/with SOC. The parity of wavefunction at high-symmetry points $\Gamma$ is showed by the minus and plus signs. Minus (plus) sign means odd (even) parity. (d) Band gap at $\Gamma$ calculated by artificially changing the strength of SOC. Band gap closing happens at artificial SOC strength around 81\%. (e) Wannier charge centers along $k_2$ on the $k_3$=0 plane. (f), (g) The calculated surface states of the AFM TI MBT on the (001) and (100) faces, respectively. The inset of (f) shows gapped surface states on the (001) face clearly, however,the inset of (g) shows no apparent gap on the (100) face.}
\end{figure}

\begin{figure}
\includegraphics[width=\linewidth]{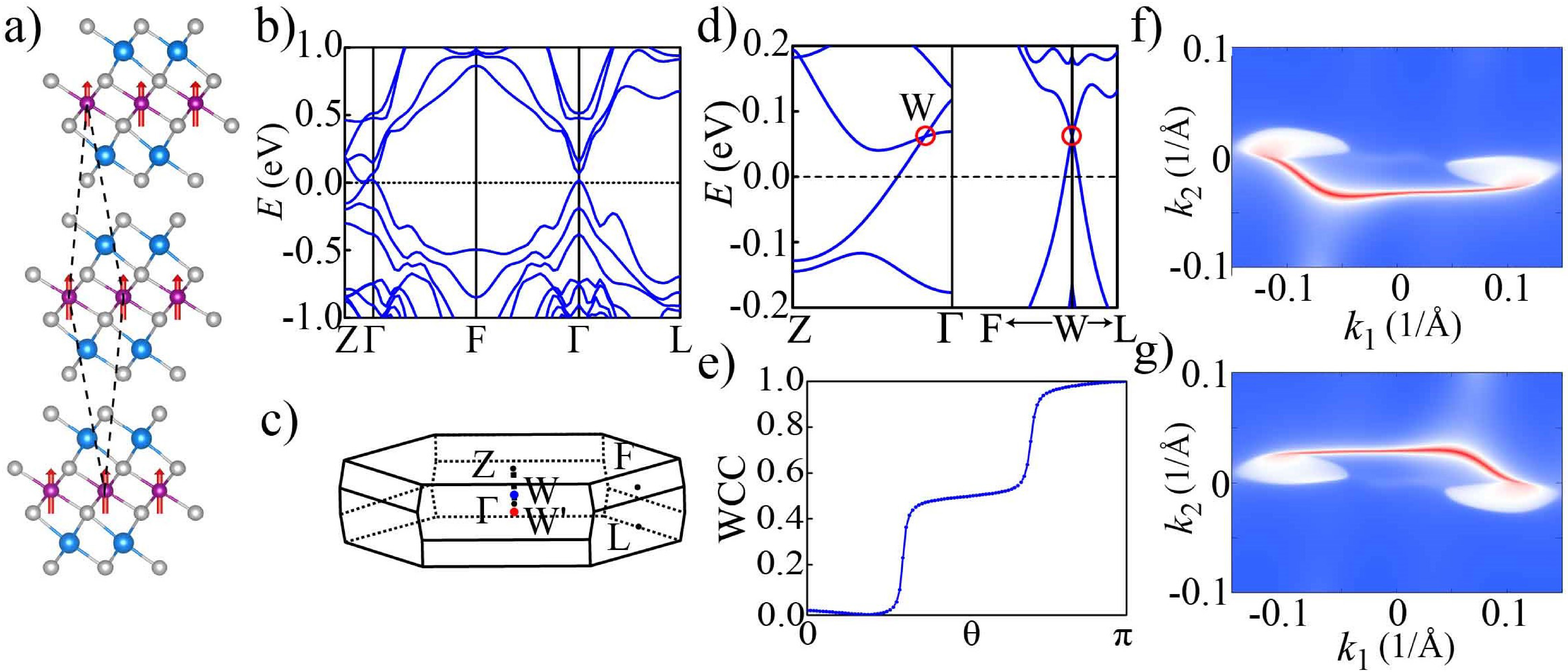}
\caption{FM MBT. (a) Atomic structure of FM MBT. Dashed rhombus represents the unit cell. (b) Band structure of FM MBT. (c) Brillouin zone of FM MBT. The red and blue points represent a pair of type-II Weyl points W and W$'$. (d) Zoom-in band structure around Weyl point W from different directions. (e) Motion of the sum of WCCs on the sphere around Type-II Weyl point W.  (f, g)  Fermi arcs on the (100) and ($\bar{1}00$) terminations. Fermi energy is fixed at the energy level of Weyl point W.}	
\end{figure}

\begin{figure}
\includegraphics[width=\linewidth]{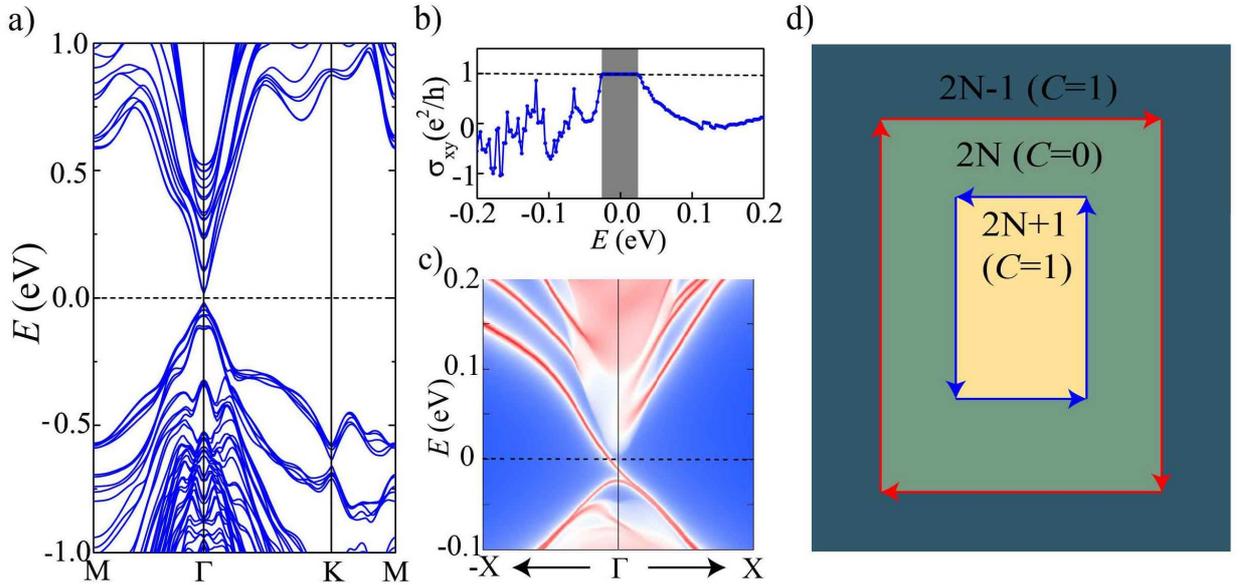}
\caption{MBT thin films. (a) Band structure, (b) Hall conductance $\sigma_{xy}$ as a function of energy and (c) edge states of the 5-layer MBT. (d) A schematic diagram showing dissipationless edge channels on step edges of MBT.}	
\end{figure}


\end{document}